\documentclass[]{emulateapj}
\usepackage{amsmath,float,xcolor}
\usepackage{natbib}
\usepackage{epstopdf}
\newcommand{\Lya}{Lyman $\alpha$\ }
\newcommand{\fesc}{$f_{\textrm{esc}}$}
\newcommand{\fc}{$f_{\textrm{cov}}$}

\newcommand{\msun}{$M_{\odot}$ }
\newcommand{\I}{\footnotesize{I}\normalsize}
\newcommand{\II}{\footnotesize{II}\normalsize}
\newcommand{\III}{\footnotesize{III}\normalsize}

\newcommand{\angstrom}{\text{\normalfont\AA}}
\shorttitle{Covering Fraction}
\shortauthors{Leethochawalit et al.}
\submitted{}

\begin{document}
\title{Absorption Line Spectroscopy of Gravitationally-Lensed Galaxies: Further Constraints on the Escape Fraction of Ionizing Photons at High Redshift}
\author{Nicha Leethochawalit\altaffilmark{1}, Tucker A. Jones\altaffilmark{2,3}, Richard S. Ellis\altaffilmark{1,4,5}, Daniel P. Stark\altaffilmark{6}, and Adi Zitrin\altaffilmark{1,3}} 
\altaffiltext{1}{Cahill Center for Astronomy and Astrophysics, California Institute of Technology, MS 249-17, Pasadena, CA 91125}
\altaffiltext{2}{Institute for Astronomy, University of Hawaii, 2680 Woodlawn Drive, Honolulu, HI 96822}
\altaffiltext{3}{Hubble Fellow}
\altaffiltext{4}{European Southern Observatory, 85748 Garching bei M\"unchen, Germany}
\altaffiltext{5}{University College London, London WC1E 6BT, UK}
\altaffiltext{6}{Department of Astronomy, University of Arizona, Tucson, AZ 85721}

\begin{abstract}
The fraction of ionizing photons escaping from high-redshift star-forming galaxies remains a key obstacle in evaluating whether galaxies were the primary agents of cosmic reionization. We previously proposed using the covering fraction of low-ionization gas, measured via deep absorption line spectroscopy, as a proxy.  We now present a significant update, sampling seven gravitationally-lensed sources at $4<z<5$. We show that the absorbing gas in our sources is spatially inhomogeneous with a median covering fraction of 66\%. Correcting for reddening according to a dust-in-cloud model, this implies an estimated absolute escape fraction of  $\simeq19\pm6$\%. With possible biases and uncertainties, collectively we find the average escape fraction could be reduced to no less than 11\%, excluding the effect of spatial variations. For one of our lensed sources, we have sufficient signal/noise to demonstrate the presence of such spatial variations and scatter in its dependence on the Ly$\alpha$ equivalent width, consistent with recent simulations. If this source is typical, our lower limit to the escape fraction could be reduced by a further factor $\simeq$2. Across our sample, we find a modest anti-correlation between the inferred escape fraction and the local star formation rate, consistent with a time delay between a burst and leaking Lyman continuum photons. Our analysis demonstrates considerable variations in the escape fraction consistent with being governed by the small-scale behavior of star-forming regions, whose activities fluctuate over short timescales. This supports the suggestion that the escape fraction may increase toward the reionization era when star formation becomes more energetic and burst-like.

\end{abstract}


\section{Introduction}
Considerable observational progress is being made in charting when cosmic reionization occurred. The most recent values of $\tau=0.058\pm0.012$, the optical depth of electron scattering of cosmic microwave 
background (CMB) photons, reported by the Planck consortium \citep{PlanckXIII15,PlanckXLVII16}, implies the average redshift at which reionization occurred lies between $z=7.8$ and 8.8. This represents a
significant revision of the picture inferred from earlier CMB data and indicates reionization was both more rapid and occurred more recently than earlier imagined. Combining the CMB data with line of sight measures
of the opacity of the intergalactic medium (IGM) obtained via spectroscopy of high redshift quasars and gamma ray bursts \citep{Chornock14} and the fraction of Lyman break galaxies whose Ly$\alpha$ emission is detectable \citep{Schenker14}, indicates reionization began shortly before $z\simeq10$ and was largely complete by $z\simeq6$.

Late reionization is particularly important in strengthening the view that star-forming galaxies were the primary ionizing agent. The abundance and luminosity distribution of $6<z<10$ galaxies is increasingly well-determined from deep imaging in blank and cluster lens fields using Hubble Space Telescope's (HST) near-infrared camera WFC3/IR\citep{Oesch15, Finkelstein15, McLeod16, Atek15}. \citet{Robertson15} present a model showing that, depending on their ionizing output,  the demographics of the $z>6$ galaxy population can match the Planck $\tau$ values by a redshift $z\simeq10$. There are two key uncertainties in the ionizing output. The first is an efficiency factor, $\xi_{ion}$, necessary to convert the (observed) UV luminosity to the production rate of Lyman-continuum photons; this depends on the nature of the stellar population and the presence or otherwise of AGN. Reasonable progress has been made in estimating typical $\xi_{ion}$ values for high redshift galaxies from recombination or other emission line measures
\citep{Bouwens15,Stark15,Stark16}. The second uncertainty is the fraction, $f_{esc}$, of the ionizing photons that escape into the IGM due to the porosity of neutral hydrogen in a typical galaxy. For reasonable values of $\xi_{ion}$, \citet{Robertson15}, \citet{Bouwens15} and \citet{Khaire16} require an absolute escape fraction, \fesc$_{\textrm{,abs}}$\footnote{\label{note1}See Section \ref{sec:def} for the definition of \fesc$_{\textrm{,abs}}$ and \fesc$_{\textrm{,rel}}$.}, of up to 20\% to account for the Planck value of $\tau$ whereas \citet{Atek15} and \citet{Mitra15} estimated \fesc$_{\textrm{,abs}}\approx10-15\%$. 

The most direct way to measure the escape fraction is to detect the leaking Lyman continuum (LyC) photons through spectroscopy or appropriate rest-frame imaging. In the local universe, non-zero spectroscopic LyC measures are only available for six galaxies. Most have \fesc$_{\textrm{,abs}}<5\%$ \citep{Leitet11,Leitet13,Borthakur14,Leitherer16}, while one has \fesc$_{\textrm{,abs}}\sim 8\%$ \citep{Izotov16}. Several surveys have attempted direct detection at $z\sim2-3$ with a similarly low success rate - a few detections out of hundreds examined \citep[e.g.][]{Vanzella15,Vanzella16,Mostardi15}. Even when significant detections are apparent, they usually reflect contaminating radiation from foreground galaxies  that can easily camouflage as substructures of the clumpy high-redshift galaxies\citep[e.g.][]{Siana15,Mostardi15}. So far, only three convincing detections have been reported: \fesc$_{\textrm{,rel}}\simeq42\%$, \fesc$_{\textrm{,rel}}>75\%$ and \fesc$_{\textrm{,rel}}>50\%$\textsuperscript{\ref{note1}} \citep{Mostardi15,Vanzella16, Shapley16}. With an assumption of some attenuation, \citet{Mostardi15} translates the measured relative escape fraction to \fesc$_{\textrm{,abs}}=14\%-19\%$. Though higher than local measurements, the measured escape fractions at $z\sim3-4$ are still relatively lower than expected considering that only the high tail of the  \fesc   distribution can be readily detected using these techniques. The relatively rare LyC leakers at $z\lesssim3$ tend to be compact, less massive, efficiently star-forming systems with high [O III]/[O II] emission line ratios \citep{Leitet13,Alexandroff15,Izotov16}. \citet{Nakajima14} present a model whereby the high [O III]/[O II] line ratio seen in many intense Ly$\alpha$ emitters can be understood if the HII regions are density- rather than ionization-bound. In this case, although a rare population locally, if such [O III]-intense systems are more common at high redshift\citep[e.g.][]{Borsani15,Smit14}, conceivably the mean escape fraction in the reionization era is significant \citep[e.g.][]{Faisst16}.
 
 Numerical simulations show mixed results regarding escape fraction and its evolution with redshift. Hydrodynamic simulations yield escape fractions ranging from smaller than a few percent \citep[e.g.][]{Wise14,Paardekooper15,Ma15} to $\sim10\%$ \citep[e.g.][]{Kimm14} in galaxies with halo masses $M\gtrsim10^9$\msun with a strong anti-correlation between escape fraction and halo mass. \citet{Ma16} show recently that inclusion of binary star evolution can boost the escape fraction at $z\sim6$ to be as high as $\sim20\%$. Most simulations find little or no evolution in the escape fraction but rather a temporary fluctuation with different delay times after bursts of star formation \citep[e.g.][]{Kimm14}.

 While observational progress will eventually be made in securing robust measures of the escape fractions for large samples at $z\lesssim3$, the above discussion illustrates how evolution in the galaxy population necessitates measures of $f_{esc}$ for sources in the reionization era. Unfortunately, the direct methods discussed above are not practical for galaxies with $z\gtrsim3$ due to the increased IGM absorption. Only a modest amount of foreground neutral hydrogen can completely absorb the LyC radiation. Therefore, even with the full armory of the James Webb Space Telescope (JWST) and next generation large ground-based telescopes, indirect methods are the only route to estimating \fesc\ and hence understanding the role of galaxies in this final piece of cosmic history. \citet{Zackrisson13} have proposed a method for estimating the LyC leakage using a combination of the UV spectral energy distribution (which constrains the stellar population) and the strength of a Balmer line such as H$\beta$ through recombination theory. For $z>6$ galaxies, this requires spectroscopic access to wavelengths beyond 2 $\mu$m and thus must await JWST.
 
A more immediate indirect method is to constrain the {\it covering fraction} $f_{cov}$ of low ionization gas from absorption line spectroscopy. As discussed in \S4 this provides a valuable proxy for the escape fraction.
The technique was first applied locally with promising results. \citet{Heckman11} studied 26 star-forming galaxies correlating the residual absorption signals they found from low ionization species with FUSE spectroscopy below the Lyman limit and the strength of recombination lines. In our earlier work \citep{Jones12} we first applied the absorption line method to a stacked spectrum derived from 81 $z\simeq$4 Lyman break galaxies (LBGs). In comparison with a similar stack at $z\simeq3$ \citep{Shapley03}, the covering fraction was reduced implying an increased porosity of neutral gas and a higher $f_{esc}$. A further comprehensive analysis of stacked spectra in the context of the escape fraction for 121 $z\simeq$3 galaxies at $z\simeq$3 has recently been presented by \citet{Reddy16}.

However, our experience with stacked spectra has revealed limitations in their utility for the present purpose. At the spectral resolution available in our large scale redshift survey at $z\simeq4-5$\citep{Jones12}, the weakened absorption found at $z\simeq$4-5 could equally reflect evolution in the outflow kinematics. By its very nature as a composite, the stacked spectrum must average over of a range of absorption line kinematic profiles, confusing any interpretation of the spectra in terms of opacity alone. Clearly it is preferable to study the joint kinematics and absorption line depths in suitably bright individual galaxies. Although a challenging task, in \citet{Jones13} we secured higher resolution absorption line spectra of 3 {\it gravitationally-lensed} $z\simeq$4 galaxies and, taking into account different kinematic profiles, confirmed the original redshift dependence mentioned in \citet{Jones12} We also found a lower covering fraction for those galaxies with intense Ly$\alpha$ emission. Although one expects $f_{esc} < 1 - f_{cov}$ and thus the method can strictly only provide an upper limit \citep{Jones13,Vasei16}, our goal here is to enlarge the sample of lensed galaxies, both in size and redshift range, so we can strengthen evidence for an evolutionary trend as well as improve our understanding of the behavior of $f_{esc}$ with other physical properties of high redshift galaxies.

In this paper, we analyze high signal to noise absorption line spectra for a further 4 lensed high redshift galaxies, more than doubling the sample in \citet{Jones13} and extending the redshift range to $z\simeq5$. This allows us to better examine the evolutionary trend in $f_{cov}$ as well as correlations with other parameters. Moreover, by studying lensed galaxies with strong gravitational magnification, we demonstrate that resolved spectroscopy allows to examine how the covering fraction varies {\it across a galaxy} as a function of its kinematics and spatially-dependent star-formation rate. 

A plan of the paper follows. We present our new data in \S2. The basic properties of our sample are then determined in \S3, including the lens models essential for recovering the intrinsic star formation rates and luminosities. We introduce two methods for deriving the escape fractions in \S4 and examine the trends derived for the total sample (including that in \citealt{Jones13}) with redshift and the Ly$\alpha$ equivalent width. We also discuss the spatially-resolved data for the $z=4.92$ galaxy lensed by the cluster MS1358+62. We summarize our results in \S5.

\section{Data}
For the present study we selected four gravitationally-lensed galaxies from the literature using the following criteria: (i) spectroscopic redshift $z>3$, (ii) total apparent magnitude $I_{AB}<23$ and (iii) the availability of a reasonable lens model to enable the magnification to be determined. In conjunction with the earlier study by \citet{Jones13}, the provides a sample of 7 lensed galaxies in the redshift range $4<z<5$.

\begin{figure*}
  \centering
  \includegraphics[width=0.3\textwidth]{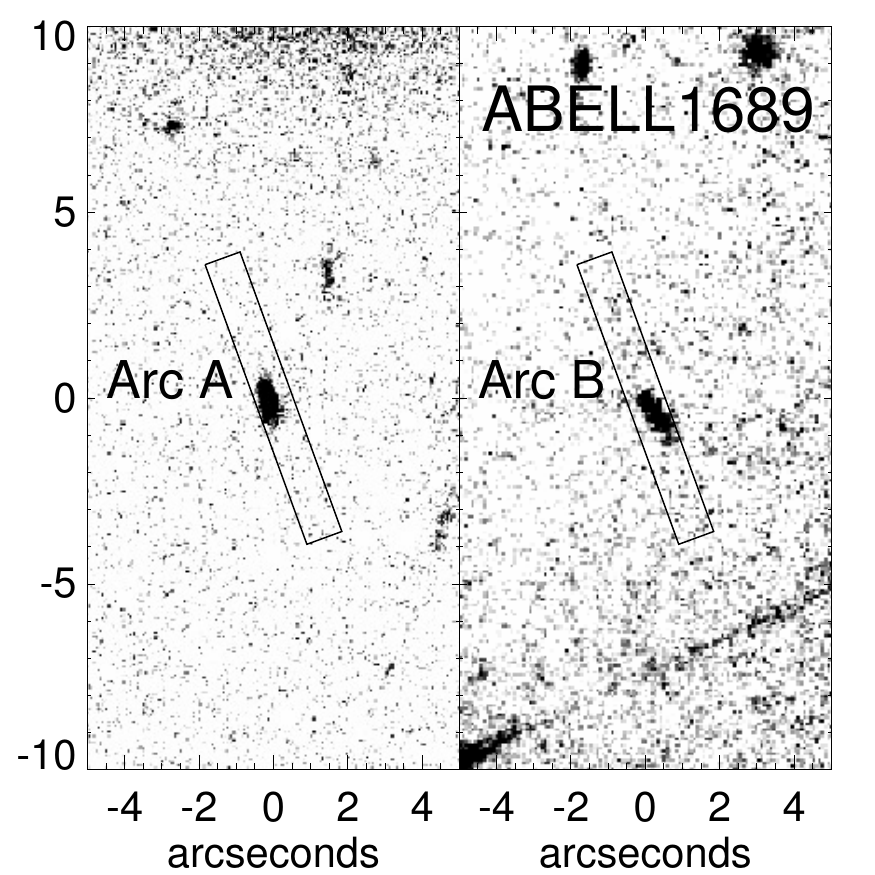}
  \includegraphics[width=0.3\textwidth]{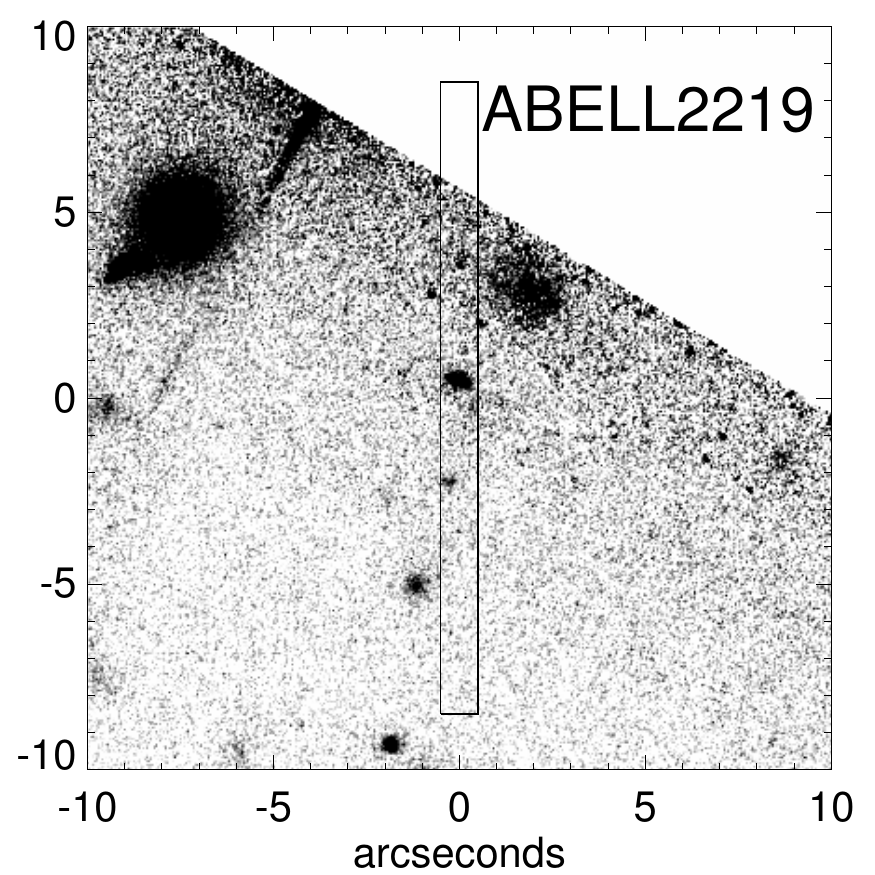}\\
  \includegraphics[width=0.3\textwidth]{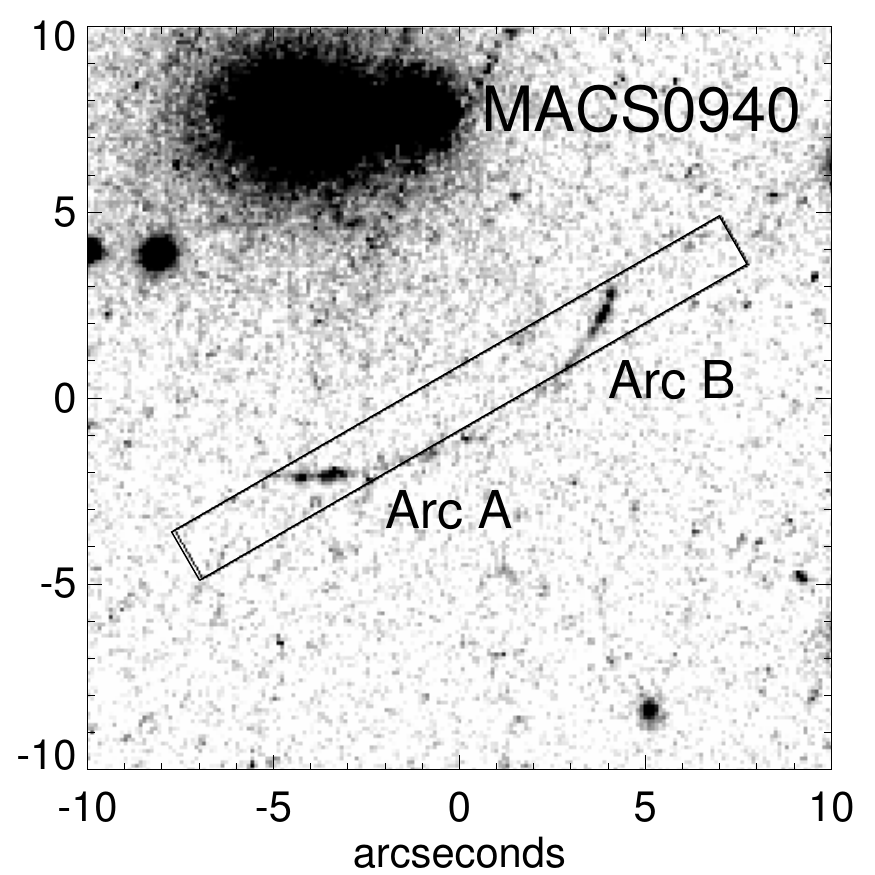}
  \includegraphics[width=0.3\textwidth]{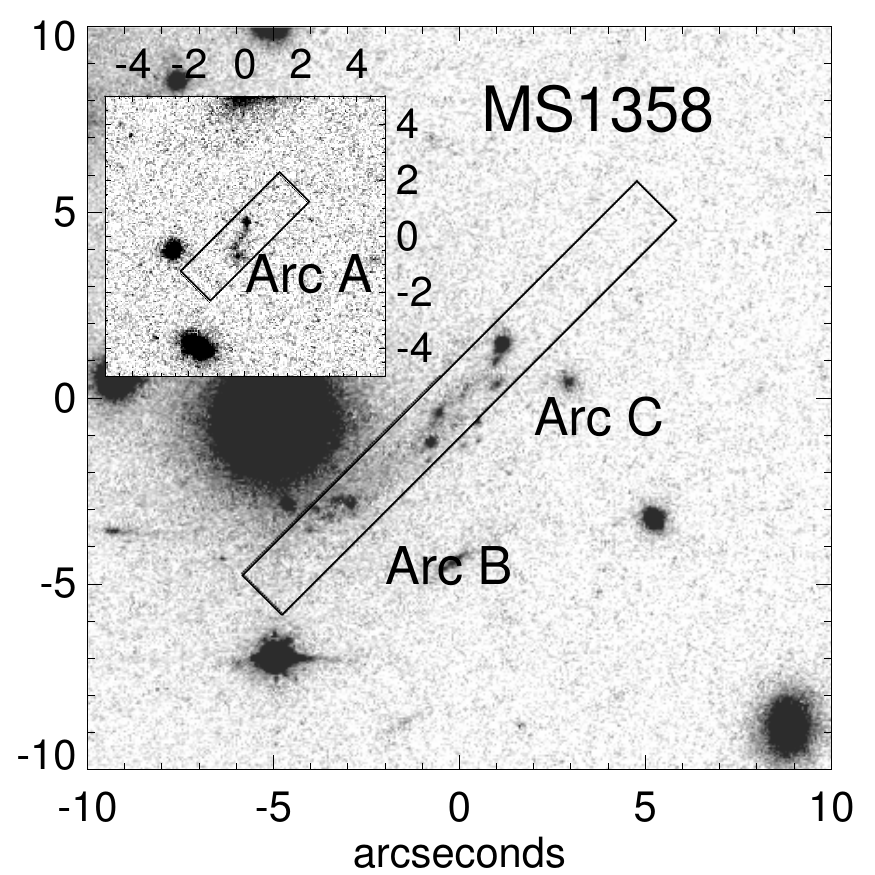}
\caption{Spectroscopic configuration for the four new arcs. The DEIMOS slit positions are overlaid on HST images for (left to right) Abell1689, Abell2219, MACS0940 and MS1358. North is up and east is to the left. The instrument/filter are ACS/F850LP for Abell 1689,  Abell 2219 and MS1358, and WFPC2/F606W for MACS0940.}
\label{fig:slit}
\end{figure*}

We observed each of the four new galaxies with the DEIMOS multi-slit spectrograph mounted on the Keck II telescope. Following the approach of \citet{Jones13}, we used a 1200 line mm$^{-1}$ grating yielding a spectral resolution of 1.8 \AA\ determined from atmospheric emission lines which corresponds to a velocity resolution of 70 km sec$^{-1}$ (see details of the observations record in Table \ref{table:observation}). This grating choice allow us to sample rest-frame wavelengths from 1200 to 1650 \AA\ ensuring good coverage of \Lya, Si\II$\lambda1260$, O\I$\lambda1302+$Si\II$\lambda1304$, C\II$\lambda1334$, and Si\II$\lambda1527$. For multiply-imaged galaxies, where possible we arranged the multi-slits to cover as many images as practical. The final slit placements are overlaid on HST images of each target in Figure \ref{fig:slit}. 

\begin{figure*}
  \centering
  \includegraphics[width=0.65\textwidth]{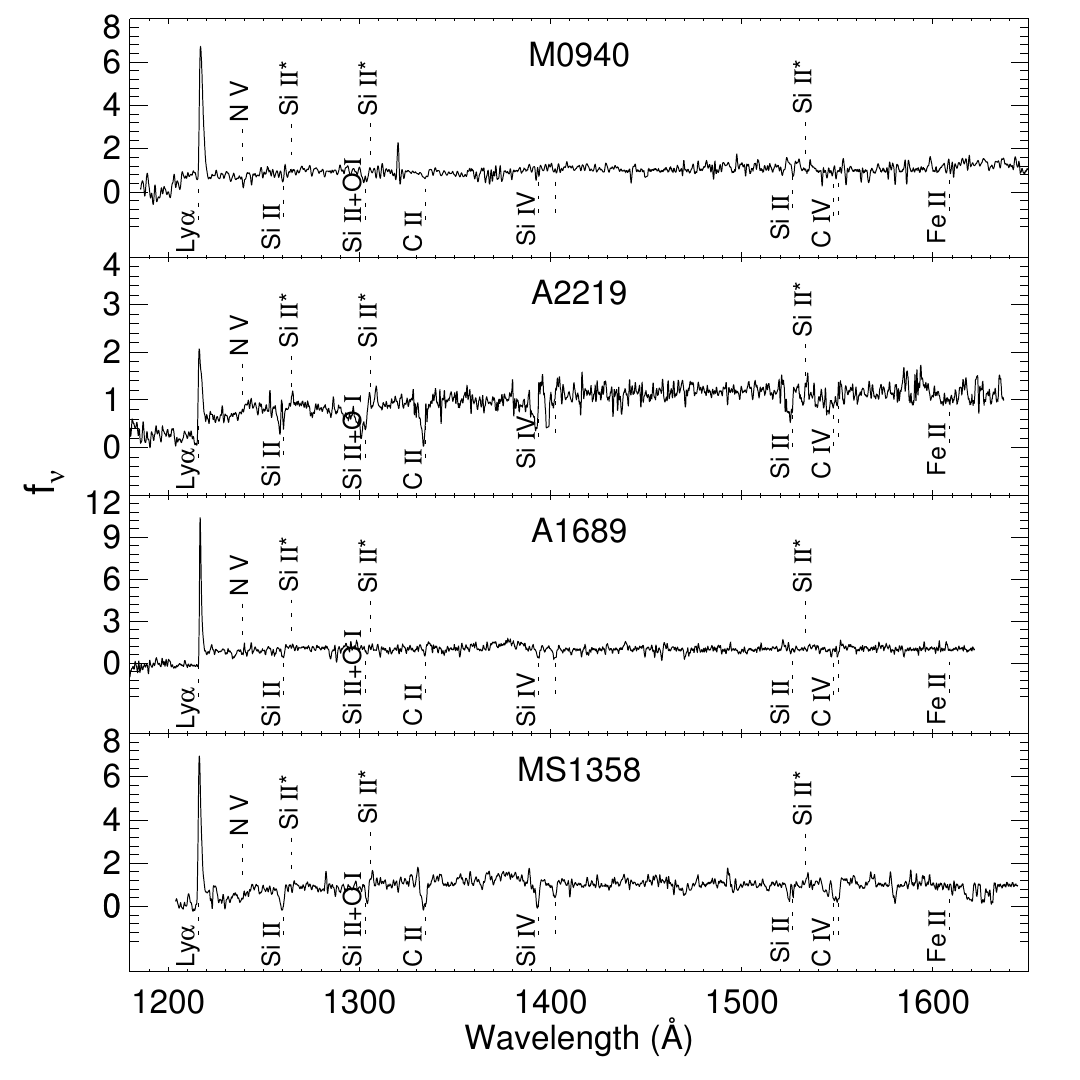}
\caption{Combined absorption line spectra for the four arcs with the low ionization lines marked.}
\label{fig:allspec}
\end{figure*}

We reduced the DEIMOS spectra using code associated with the SPEC2D pipeline. Because our galaxies are lensed with extended non-Gaussian profiles, we used a boxcar extraction with specified full width half maxima (FWHM) to extract the 1D spectra from the 2D data. Each spectrum was corrected for telluric absorption line using spectra of standard stars and smoothed with the spectral resolution at each wavelength measured from atmospheric emission lines. For absorption line measurements, we spline fitted the continuum and normallized each spectrum. If the galaxy was observed with multiple masks (in the case of MS1358) or had multiple images (MACS0940), we combined the continuum-normallized 1D-spectra using a weighted arithmetic mean. For MS1358, we excluded Arc B in the final combination both to prevent contamination from a foreground lensing galaxy and since some sub-components in the lensed image lay outside the slit. For star formation rate measurements, we calibrated each telluric-corrected spectrum using standard stars observed at the end of each night for an overall throughput correction and flux calibrated using HST photometry in that imaging band which overlaps our DEIMOS wavelength coverage. The spectra of each galaxy were finally combined using weighted arithmetic mean. The final spectra are shown in Figure \ref{fig:allspec}; the typical signal-to-noise is $\sim10$ per spectral resolution element. 

\begin{deluxetable*}{cccccccc} 
\tabletypesize{\footnotesize} 
\tablecolumns{9} 
\tablewidth{0pt} 
\tablecaption{Observation data \label{table:observation}} 
\tablehead{\colhead{ID}&\colhead{Arc} & \colhead{RA} & \colhead{DEC} & \colhead{$z_{\textrm{ISM}}$} & \colhead{$z_{\textrm{Ly}\alpha}$}   & \colhead{Date} &  \colhead{t$_{\text{exp}}$(hr)}}   
\startdata
MACS 0940 & ArcAB & 09:40:53 & +07:44:18 &$4.031\pm0.001$ & $4.034$   & 03/15 &       5.5\\
Abell 2219&      & 16:40:16 & +46:43:59 &$4.450\pm0.002$ & $4.451$   & 05/15 & 6.2    \\
Abell 1689& Arc A& 13:11:25 & -01:20:52 &$4.868\pm0.001$ & $4.875$   & 05/15 & 4.9 \\
          & Arc B& 13:11:31 & -01:20:14 &&&&\\
MS 1358   & ArcA & 13:59:55 & +62:31:04 &$4.928 \pm 0.002$ & 4.927  & 03/15\&02/16 &       7.5 \\
          & ArcBC & 13:59:49 & +62:30:48&                        &       &       \\
\cutinhead{\emph{Objects published in \citet{Jones13}}}
Abell2390\_H3a &      & 21:53:34 & +17:42:03 &$4.043\pm0.002$ &  & 10/11,06/12 & 10.8    \\
Abell 2390\_H5b&      & 21:53:35 & +17:41:33 &$4.0448\pm0.0003$ &  & 10/11,06/12 & 10.8    \\
1621+0607      &      & 16:21:33 & +06:07:05 &$4.1278\pm0.0004$ &  & 10/11,06/12 & 5.7  
\enddata 
\end{deluxetable*}

\section{Derived Properties}
\label{sec:props}

We first derive the basic physical properties of our sample including the magnification for each lensed galaxy determined from a cluster mass model. These are summarized in Table 2.

\subsection{Systemic Redshift}

The systemic redshift is an important measure since it enables us to examine the spectra for the presence of outflowing gas. Accurate systemic redshifts are usually obtained from stellar absorption lines or the nebular emission lines arising from HII regions. For our sample, a redshift from nebular [O II] 3727 \AA\ emission is only available for MS1358 \citep{Swinbank09}. However, since our spectra primarily target low-ionization absorption lines and \Lya emission, it is more practical to use these to estimate the systemic redshift (which is in reasonable agreement with that derived from nebular emission for MS1358). \citet{Jones12} determined an offset of  $\Delta v_{IS}$=190 km s$^{-1}$ between the systemic redshift ($z_{sys}$) determined from the C\III$\lambda1176$ stellar photospheric line to that measuring using the low-ionization absorption lines ($z_{IS}$) using a composite spectrum of 81 LBGs at $z\sim4$. Following \citet{Jones13}, we apply these velocity offsets to the average weighted centroid of low-ionization absorption lines to determine the systemic redshift of each galaxy. We used all prominent low-ionization absorption lines available namely Si\II$\lambda1260$, O\I$\lambda1302$, Si\II$\lambda1304$, C\II$\lambda1334$, and Si\II$\lambda1527$. The uncertainty is a square sum of the standard deviation of redshifts measured from different absorption lines and the rms scatter $\sim125$ km s$^{-1}$ between redshifts derived from absorption lines and those from nebular emission \citep{Steidel10}. Together with the redshift of Ly$\alpha$, the results are listed in Table \ref{table:observation}.   


\subsection{Star Formation Rates}

We measure the star formation rates (SFR) of our lensed targets using the extinction-corrected 1600 \angstrom\  flux from the observed DEIMOS spectra. To account for dust extinction, we measure the UV continuum slope($\beta$)\footnote{Ultraviolet continua are usually parametrized via a power-low index, $\beta$, where $f_\lambda \propto \lambda^\beta$ \citep{Meurer99}} of the spectra via a linear fit to magnitude as a function of logarithmic wavelength over 1250-1650 \angstrom\ in rest frame. It is not possible to estimate the UV slope using photometric data for 3 of our recently-observed sample due to the absence of one or more of the key photometric bands. However, for MS1358 we can compare our spectroscopically-measured slope with that derived from available HST/ACS F850LP and HST/WFC3 F110W data. We find reasonable agreement given the uncertainties arising from the faint and extended nature of this object. Our UV slopes range from $\beta \sim -1.4$ to $-2.8$ in good agreement with those determined for large samples of LBGs at similar redshifts \citep[e.g.][]{Bouwens09, Castellano12}. The \citet{Calzetti00} extinction law is applied to translate the UV slope to the extinction at 1600 \angstrom. The SFR is then calculated from the extinction-corrected 1600 \angstrom\ flux according to the relation given by \citet{Kennicutt98}, in which a Salpeter \citeyearpar{Salpeter55} initial mass function was assumed. Finally, we used the magnification from the lens models described in Section \ref{sec:magnification} to derive the intrinsic SFR. For the three lensed galaxies from \citet{Jones13}, we adopt a similar procedure except the UV slopes were determined from the appropriate rest-frame UV HST photometry. For Abell 2390\_H3a, the UV slope is calculated from HST/WFPC2 F814W and Keck/NIRC H band photometry reported in \citet{Bunker00}. For Abell 2390\_H5b and J1621+0647, the UV slopes are calculated from HST/WFPC2 F814W, HST/WFC3 F125W, and F110W images. 

For the arc in the M1358+62 cluster, we can compare our derived SFR to that estimated from other methods in the literature. For the north-west arc (Arc C in Figure \ref{fig:slit}), we measure 
$\textrm{SFR}_{\textrm{UV}}=594\pm202$ M$_\odot \textrm{yr}^{-1}$ (uncorrected for lensing). The result is consistent with that derived from an [OII] emission line flux with no reddening correction at $\textrm{SFR}_{\textrm{O[II]}}=525\pm55$ M$_\odot \textrm{yr}^{-1}$ \citep{Swinbank09}. Similarily, the unlensed SFRs for Abell 2390\_H3a and Abell 2390\_H5b of $95\pm52$ and $13\pm5 M_\odot$ yr$^{-1}$ are consistent with estimates reported in \citet{Pello99} using SED-model fittings of $\sim60$ and $\sim10 M_\odot$ yr$^{-1}$ respectively.
 
\begin{deluxetable*}{cccccc} 
\tabletypesize{\footnotesize} 
\tablecolumns{6} 
\tablewidth{0pt} 
\tablecaption{Extinction and Star Formation Rates \label{table:SFR}}
\tablehead{\colhead{Cluster} & \colhead{Arc} & \colhead{Magnification} & \colhead{UV slope} & \colhead{SFR}& \colhead{$\overline{\textrm{SFR}}$}\\
\colhead{}&\colhead{}&\colhead{($\mu$)}&\colhead{($\beta$)}&\colhead{(M$_\odot$ yr$^{-1}$)}&\colhead{(M$_\odot$ yr$^{-1}$)}}
\startdata
MACS 0940 &arcA&$     182\pm411$&$     -1.49\pm  0.11$&$           1\pm           3$&$      2\pm  3$\\
          &arcB&$      38\pm 19$&$     -1.35\pm  0.26$&$           9\pm          10$\\
Abell 2219&    &$       1.6\pm  0.3$&$     -1.76\pm  0.09$&$          89\pm          17$&\\
Abell 1689&arcA&$       5.3\pm  1.1$&$     -1.76\pm  0.07$&$          46\pm           9$&$     47\pm  9$\\
          &arcB&$       2.2\pm  0.4$&$     -1.92\pm  0.21$&$          44\pm          45$\\
         
MS 1358   &arcA&$       2.9\pm  0.6$&$     -1.60\pm  0.11$&$          74\pm          14$&$     73\pm 14$\\
          &arcC&$      12.1\pm  6.1$&$     -1.67\pm  0.04$&$          66\pm          33$\\
          
Abell 2390\_H3a &   &$       19\pm  10$  &$     -1.75\pm  0.10$&$          95\pm          52$&\\
Abell 2390\_H5b &   &$       6.5\pm  1.3$&$     -2.81\pm  0.15$&$          13\pm          5$&\\
1621+0607  &   &$       38\pm  19$&$     -1.9\pm  0.2$&$          31\pm         20$&        \\

\cutinhead{Clumps in MS1358}
MS1358&C0&$      10.0\pm  2.0$&$     -1.74\pm  0.07$&$          33\pm           6$\\
MS1358&C1,2,3&$      11.5\pm  5.8$&$     -1.89\pm  0.08$&$          15\pm           7$\\
MS1358&C4&$      13.1\pm  6.6$&$     -1.99\pm  0.09$&$           6\pm           3$\\

\enddata 
\end{deluxetable*}

\subsection{Lens Models and Magnifications}
\label{sec:magnification}
Gravitational lensing enables us to directly measure covering fractions for galaxies at high redshift without the need for lens models since the magnification involved has no wavelength dependence. However to derive the intrinsic properties such as the star formation rate, an accurate mass model for the lensing cluster is required. For each galaxy cluster we constructed a mass model using the Light-Traces-Mass (LTM) method (full details in \citealt{Zitrin15} and \citealt{Zitrin09}). The model consists of a galaxy component, and a smooth dark matter (DM) component. The galaxy component is a superposition of all cluster member galaxies, where the model assumes the mass of each galaxy scales with its luminosity and can be described via an elliptical shaped power law profile. A core is often also introduced, especially for the brightest central galaxies (BCGs). The mass-to-light ratio is fixed for all cluster members except the BCG. The DM component is a smoothed version of the galaxy component, and the two components are added with a relative weight optimized in the minimization procedure. In addition, a two-component external shear is added to allow for further flexibility. We use a Markov Chain Monte Carlo (MCMC) to construct the mass model by minimizing differences between the predicted and actual position of multiple images. The multiple images and their spectroscopic redshifts were obtained from the literature i.e. \citet{Swinbank09, Zitrin11, Smith05, Gladders02, Rzepecki07}.

Using the mass model, we transfer the rest-frame UV image into the source plane and determine the magnification $\mu$ from the associated flux ratio. Uncertainties arise both from the precision of the model,
which can be estimated from the predicted versus observed position of multiple images, and systematic effects associated with the LTM model compared, for example with a model based on analytic profiles (see \citet{Zitrin15}). We adopt an uncertainty of 20\% when the magnification $\mu<10$ and 50\% if $\mu>10$. The magnification $\mu=12.1\pm6.1$ derived for MS1358 is consistent with the value of $\mu\simeq12.5$  derived by \citet{Swinbank09} offering a valuable confirmation of our methods.


\medskip
\section{Covering and Escape Fractions}
\subsection{Definitions}
\label{sec:def}

The goal of this paper is to interpret the covering fraction of low ionization gas in terms of new constraints on the escape fraction of ionizing photons in an enlarged sample of $4<z<5$ galaxies. We will introduce two methods for making the connection between the covering and escape fraction. However, first it is important to introduce some definitions. 

The most important measure of the escape fraction is that used in calculating whether galaxies are capable of driving cosmic reionization. This is the {\it absolute escape fraction} defined as: 

$$f_{\textrm{esc,abs}}\equiv \frac{F_{\textrm{LyC,out}}}{F_{\textrm{LyC,int}}}$$

that is the ratio between the Lyman continuum flux density escaping the galaxy, usually measured at 900 \angstrom, to the internal stellar flux density at the same wavelength. Most of the escape fractions reported in numerical simulations and theoretical calculations refer to this absolute value. However, observationally it is very difficult to estimate the internal stellar flux density at 900 \angstrom\ since it requires modeling the spectral energy distribution(SED) beyond the range of direct observation. Instead, a more accessible estimate is the {\it relative escape fraction} defined as:

$$f_{\textrm{esc,rel}}\equiv \frac{(f_{\textrm{LyC}}/f_{1500})_{\textrm{out}}}{(f_{\textrm{LyC}}/f_{1500})_{\textrm{int}}}=\frac{f_{\textrm{esc,abs}}}{10^{-0.4A_{1500}}}$$

Since $f_\textrm{LyC,out}$ can be additionally absorbed by neutral hydrogen in the IGM along the line of sight, we can write $f_{\textrm{esc,rel}}$ as a function of the observed Lyman continuum flux($f_{\textrm{LyC,obs}}$) as

$$f_{\textrm{esc,rel}}=\frac{(f_{\textrm{LyC}}/f_{1500})_{\textrm{obs}}}{(f_{\textrm{LyC}}/f_{1500})_{\textrm{int}}}e^{\tau_{_{HI,\textrm{IGM}}}}$$

Notwithstanding the fact that the relative escape fraction can be obtained from good observations, modeling of the SEDs is still required to determine the intrinsic ratio of the 900 \angstrom\  and 1500 \angstrom\ luminosities, which results in additional uncertainties.

\subsection{Method 1: Low-ionization Covering Fraction}
\label{sec:fullmethod}
We can estimate the covering fraction of neutral hydrogen in a galaxy by measuring the amount of non-ionizing UV radiation absorbed by metals following the approach described in \citet{Heckman11} and \citet{Jones13}. Realistically, we must assume that galaxies are covered partially by HI regions with different velocities relative to the systemic value. Neutral clouds of a particular velocity will have a covering fraction of \fc$(v)$. Photons of each low-ionization absorption line, when shifted to match the cloud velocity, will be able to escape if they travel through holes in the neutral gas or will otherwise be attenuated. The underlying principle is that low ionization gas represents a faithful tracer of neutral hydrogen, an assumption discussed in detail by \citet{Jones13}, \citet{Henry15} and \citet{Vasei16}. \citet{Trainor15} find low ionization absorption line widths correlate with the inferred escape of Lyman $\alpha$ photons, and \citet{Rivera15} find a strong correlation with 21 cm line widths which directly trace the neutral hydrogen. However, \citet{Reddy16} claim, from their recent $z\simeq$3 sample, that the outflowing gas may be metal poor in which case the true HI covering fraction could be underestimated somewhat by our technique.

The relevant radiative transfer equation is:
\begin{equation}
\frac{I(v)}{I_0} = (1-f_{\textrm{cov}}(v))+f_{\textrm{cov}}(v)e^{-\tau(v)}
\label{eq:I}
\end{equation}

where $I_0$ is the continuum level and $\tau$ is the optical depth of each absorption line:
\begin{equation}
\tau(v) = \frac{\pi e^2}{m_ec}N(v)f_{lu}\lambda_{lu}=\frac{N(v)}{3.768\times10^{14}}f_{lu}\lambda_{lu}
\label{eq:tau}
\end{equation}
where $f_{lu}$ is an oscillator strength of each transition from level $l$ to $u$ (available from the NIST Atomic Spectra Databased), $\lambda$ is the transition wavelength in \angstrom, and $N(v)$ is the ion column density in cm$^{-2}$(km s$^{-1})^{-1}$. Equation \ref{eq:tau} assumes none of the excited states are populated, i.e. $N_u/N_l\approx 0$.  Combining equations \ref{eq:I} and \ref{eq:tau}, the attenuation in the continuum normalized spectra at each velocity of an absorption line is then a function of two variables: \fc$(v)$ and column density of absorbers $N(v)$ with velocity $v$. If we have more than two absorption lines originating from the same ion, we can solve for the two variables.

This picture of a ``picket fence'' for the ISM, whereby each individual absorbing cloud forms a `picket' for each velocity bin, is consistent with that determined to be an appropriate model for local analogs of high redshift star-forming galaxies. A good example follows observations of the Ca\II H and K absorption lines in clouds around hot OB stars that resolve into several radial velocity components \citep{Adams49}. Moreover, \citet{Heckman11} found all four local Lyman Break Analogs with high residual intensities in the low-ionization absorption lines are a better match to the picket fence model compared to a uniform shell model.

\begin{figure*}
  \centering
  \includegraphics[width=0.35\textwidth]{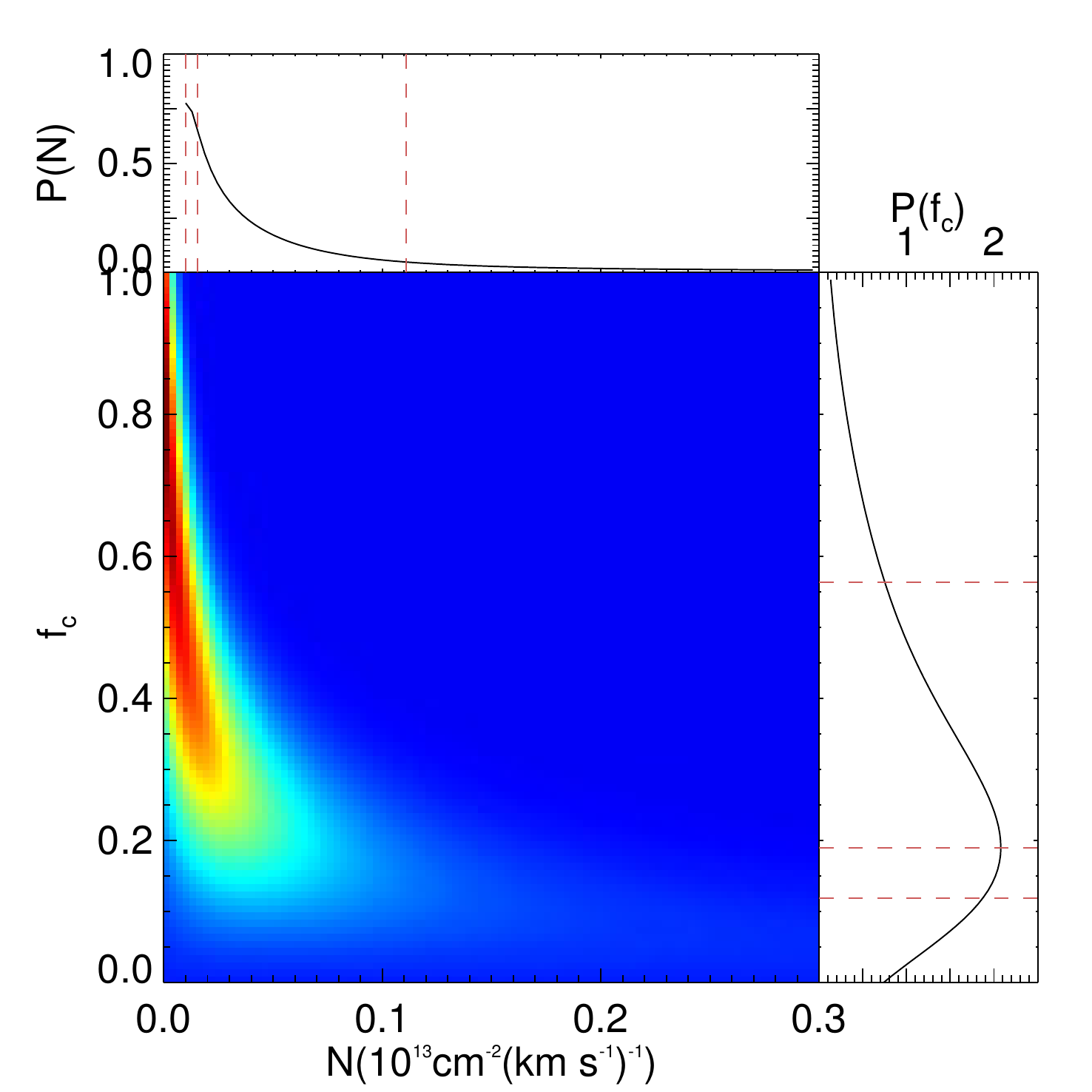}
  \includegraphics[width=0.35\textwidth]{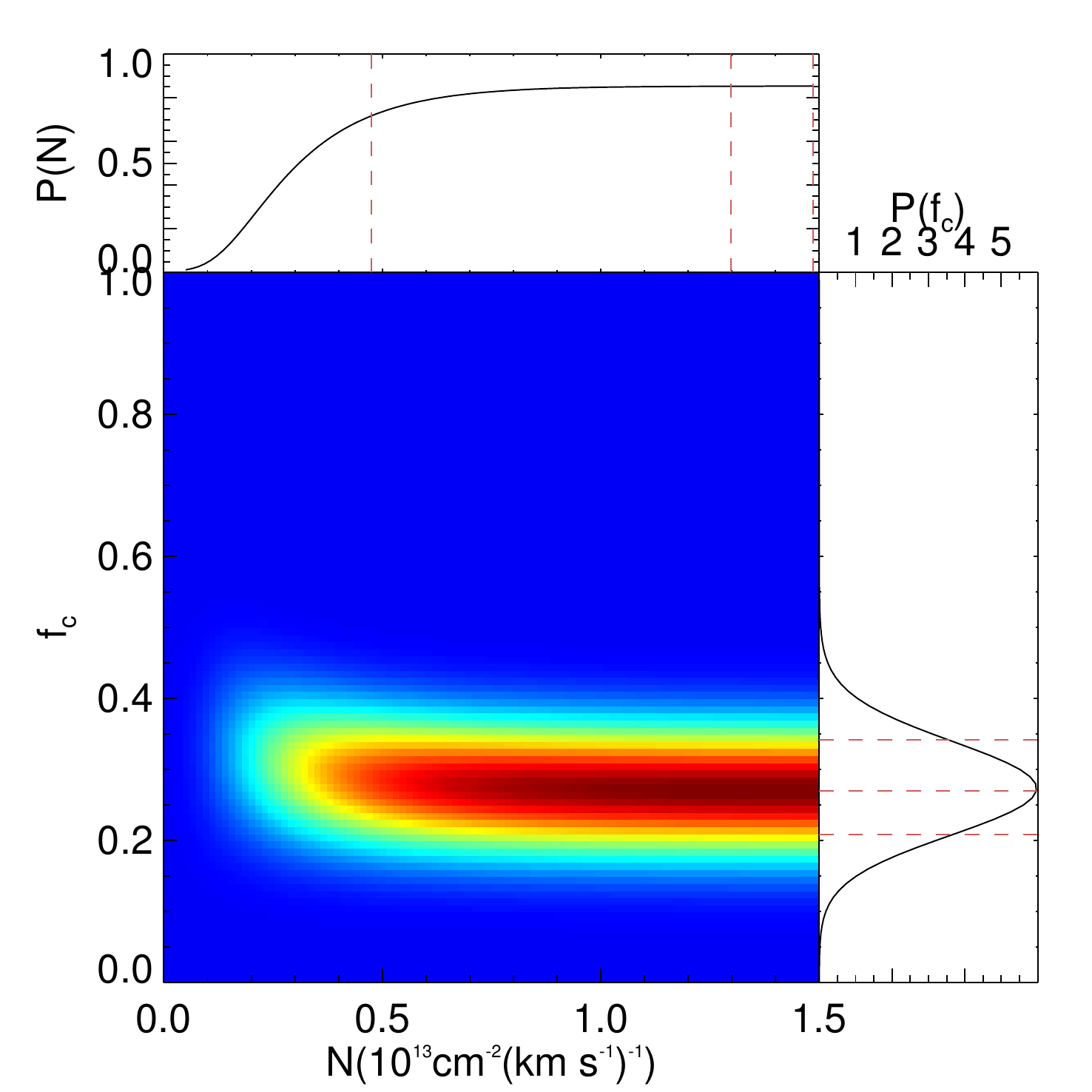}
  \caption{Posterior distributions of covering fractions (\fc) and column densities($N$) in MACS 0940 at velocity $v=-270\pm50$ km s$^{-1}$ (left) and at $v=-20\pm50$ km s$^{-1}$ (right). The likelihood are calculated from three SiII transitions at 1260, 1304 and 1526 \angstrom. The middle color plots represent 2-D posterior distribution with their marginalized 1-D posterior distributions of column densities and covering fractions projected on the top and on the right. The best values(median) and their $1\sigma$ uncertainties are shown in dashed lines. Most of the posterior distributions show optically thick absorption lines as shown in the right figure.}
\label{fig:pdf}
\end{figure*}

Our DEIMOS spectra cover three absorption line transitions of SiII at 1260, 1304 and 1526 \angstrom\  enabling us to solve for both \fc$(v)$ and $N(v)$. We bin the absorption lines into several velocity components with the width of each velocity bin chosen to be equal to the spectral resolution at that wavelength. We limit the usage of Si \II $\lambda1304$ to only those systems whose velocity $v \gtrsim -200$ km s$^{-1}$ in order to avoid contamination by O \I $\lambda1302$. We use a brute force (grid search) technique to find the best fit parameters for each velocity bin. The likelihood of each pair of parameters is calculated from the least-square residual $\chi^2=\sum(I_{\textrm{obs}}-I_{N,f_c})^2$. We adopt a prior range for \fc$(v)$ of [0,1] while the prior for $N(v)$ is adjusted so that the range at each velocity covers the whole posterior probability distribution. An exception is made is when the lines are optically thick, in which case only a lower limit on the column density can be obtained. Examples of posterior distributions for both optically thin and optically thick regimes are shown in Figure \ref{fig:pdf}.

\subsection{Method 2: Average Low-ionization Absorption Profiles}
\label{sec:simplemethod}

In our second method, we simplify the covering fraction model given in Section \ref{sec:fullmethod} by assuming that all clouds are optically thick $\tau\gg1$. Equation \ref{eq:I} then becomes
\begin{equation}
f_{\textrm{cov}}(v) = 1-\frac{I(v)}{I_0}
\label{eq:II}
\end{equation}

This enables us to include absorption lines from other species present in our spectra, including OI$\lambda1302$ and CII$\lambda1334$. We will see that the optically thick case is a reasonable assumption as results from Method 1 show that most of the SiII absorption lines are optically thick\footnote{SiII 1260, 1304, and 1526 are optically thick when N(SiII) $\gtrsim [0.02,0.31,0.18]\times 10^{13}$ cm$^{-2}$. OI 1302 and and CII1334 are optically thick when N(OI)$\gtrsim0.56 \times 10^{13}$cm$^{-2}$ and N(CII) $\gtrsim 0.22 \times 10^{12}$cm$^{-2}$ respectively.}. At each cloud velocity, we then calculate the covering fraction as an inverse-variance weighted mean of the line residuals. 
        
\subsection{Escape Fraction Constraints}

\begin{figure*}
  \centering
  \includegraphics[width=0.6\textwidth]{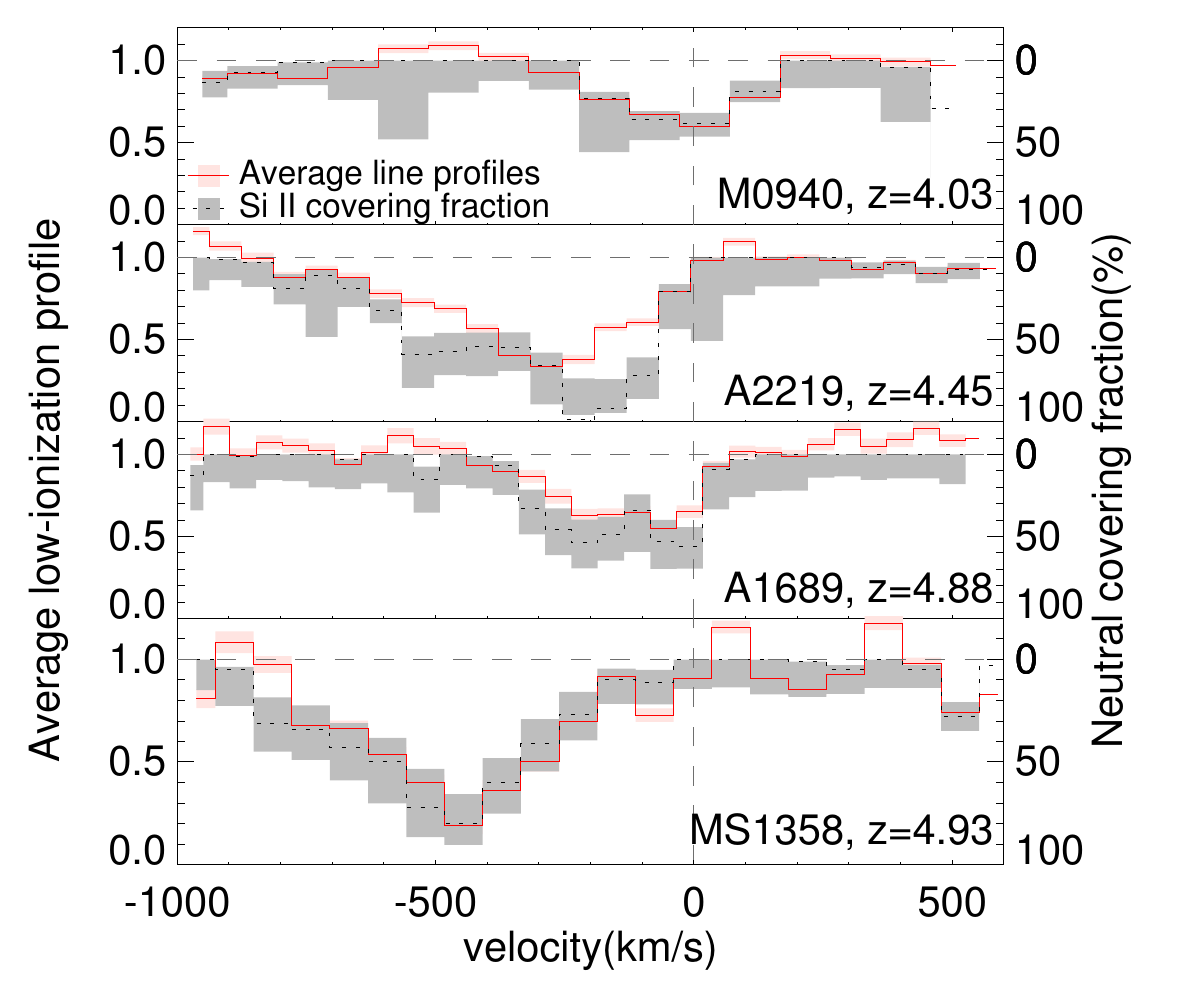}
  \caption{Average low-ionization absorption line profiles and covering fraction derived in Section \ref{sec:fullmethod} (black lines) and Section \ref{sec:simplemethod} (red lines). The shaded areas show $1\sigma$ uncertainties for each method. Each velocity bin corresponds to one spectral resolution. With no dust assumption, the covering fraction ranges from $\sim50\%-100\%$ at the central velocities.}
  \label{fig:covfrac}
\end{figure*}

The covering fractions estimated from low-ionization absorption profiles at each velocity resolution element are shown in Figure~\ref{fig:covfrac}. In general we find that the absorbing gas is spatially inhomogeneous: the absorption line profiles show only partial covering of the stellar radiation with a maximum absorption depth of $\sim$50--100\%. Formally this translates to geometric escape fractions of up to 50\%. Only two of the seven lensed galaxies, A2219 and J1621, are consistent with a complete (100\%) covering fraction at any velocity and hence a zero escape fraction. Without accounting for possible spatial variations in the covering fraction as a function of velocity, which will be examined in Section \ref{sec:spatial_variation}, we find a median value of covering fraction in our enlarged sample is $f_{cov,obs} = 66\%$. If this represents the true HI covering fraction, corresponds to $f_{esc,rel} = 34\%$. However, to translate these measures to the desired absolute escape fractions, it is necessary to account for dust extinction since these galaxies are not dust-free as indicated by the fact their UV slopes $\beta > -2$  (see Table~\ref{table:SFR}). Accounting for reddening based on the dust-in-cloud model \citep[see][]{Vasei16}, the corresponding median $f_{esc,abs}$ and weighted standard deviation of our sample becomes $19\pm6$\%.

At face value such an absolute escape fraction, if typical of sources at higher redshift, would imply that early star-forming galaxies are readily capable of maintaining reionization\citep[e.g.][]{Robertson15}. Our median value derived from low-ionization absorption profiles is consistent though somewhat larger than that inferred via recombination rate studies of the IGM using QSO absorption lines at a similar redshift, which imply $f_{esc}\sim12\%$ \citep{Becker15}. If $f_{esc}$ is indeed 12\% then we would expect an average $f_{cov,obs}\sim78$\% for our sample, accounting for reddening. It is likely that our estimate is higher than the true value due to several violations of the assumed simple ``picket fence'' model of absorption as discussed in \citet{Jones13} and \citet{Vasei16}. We now explore these possibilities quantitatively in order to estimate the degree by which our $f_{cov,obs}$ measures may overestimate the true escape fraction.

Firstly, it is possible that the non-ionizing stellar radiation could be associated with a lower covering fraction of gas than that for the ionizing starlight. This would arise if, for example, the H I and low-energy ions preferentially cover younger stars. One way to estimate the potential importance of this effect is to estimate the fraction of detected UV continuum radiation (at $\sim$1500 \AA) which arises from starlight whose photons are incapable of ionizing hydrogen. For a galaxy with a history of continuous star formation over 100 Myr, only around 10\% of the UV radiation arises from spectral types later than B \citep{Parravano03} so this cannot be a significant effect. Furthermore the Si II transitions show no evidence of a higher covering fraction at shorter wavelengths. Such an effect would result in weaker absorption of Si II $\lambda$1526 \AA\ compared to $\lambda$1304 \AA\ in an optically thick gas, which is not observed in the majority of our combined sample where both lines are free from sky contamination. We therefore conclude that a differential covering fraction of ionizing stars introduces $<10$\% systematic error on our derived values.

A second uncertainty arises from the fact that scattered photons are capable of ``filling in'' the absorption profiles, mimicking a reduced covering fraction. In the extreme case where line filling is induced from optically thin gas, the Si II 1304 and 1526 \AA\ absorption profiles would be filled by ~50\% of the flux seen in the Si II* 1309 and 1533 lines, respectively. The observed fine structure emission line fluxes therefore allow us to directly estimate the magnitude of this effect. In Abell 2219 where these emission lines are strongest, we find that scattering could only reduce the inferred covering fractions by at most 13\%. On average across the full sample, we find line filling can change our derived $f_{cov,obs}$ by at most 4\% (i.e. to $f_{cov,obs}=70$\%) which is small compared to the quoted uncertainties. The actual effect of line filling must be even less than this limit as most cases indicate optically thick absorption.

In summary, accounting for the possibilities of wavelength-dependent covering fraction and line filling discussed above, we can revise the original covering fraction to at most $f_{cov,obs} = 66 + 10 + 4 = 80$\% implying $f_{esc,abs}\sim11$\% . However, we emphasize that these corrections represent the maximum conservative limits allowed by our data, and there is no conclusive evidence that the effects discussed actually reduce $f_{esc}$ at all. Therefore the difference between our covering fraction measurement and the $f_{esc,abs}\sim12$\% implied by IGM studies, if real, can only be reconciled if these are the only important effects. 

The most challenging and remaining uncertainty arises from the possibility that absorbing gas at different velocities may cover different spatial lines of sight \citep[e.g.][]{Rivera15}. To examine this possibility, we would need to examine the {\it spatial variation} of the absorption line profiles. Normally this would currently be impractical in any source at $z>4$ but, by good fortune, this is possible in our most distant lensed source MS1358. In the next section, we show that it is likely that such spatial variations may be the largest cause of difference between our derived covering fractions and the true ionizing escape fraction.
   
\subsection{Spatial Variations in the Low Ionization Gas}
\label{sec:spatial_variation}
In the specific case of the highly-magnified target MS1358 at $z=4.93$, we can examine the covering fraction of low ionization gas in discrete sub-components of the galaxy. When the region being examined spectroscopically is physically small, the angular distribution of clouds of different velocities should be less of a concern. In addition, if a particular region has a high star-formation rate, the possibility that its stellar continuum is affected by non-ionizing radiation should also be reduced. By examining the spectroscopic variation among the star-forming clumps in MS1358, we can therefore address several of the uncertainties in the inferred escape fraction from our methods discussed above.

\begin{figure}
\includegraphics[width=0.24\textwidth]{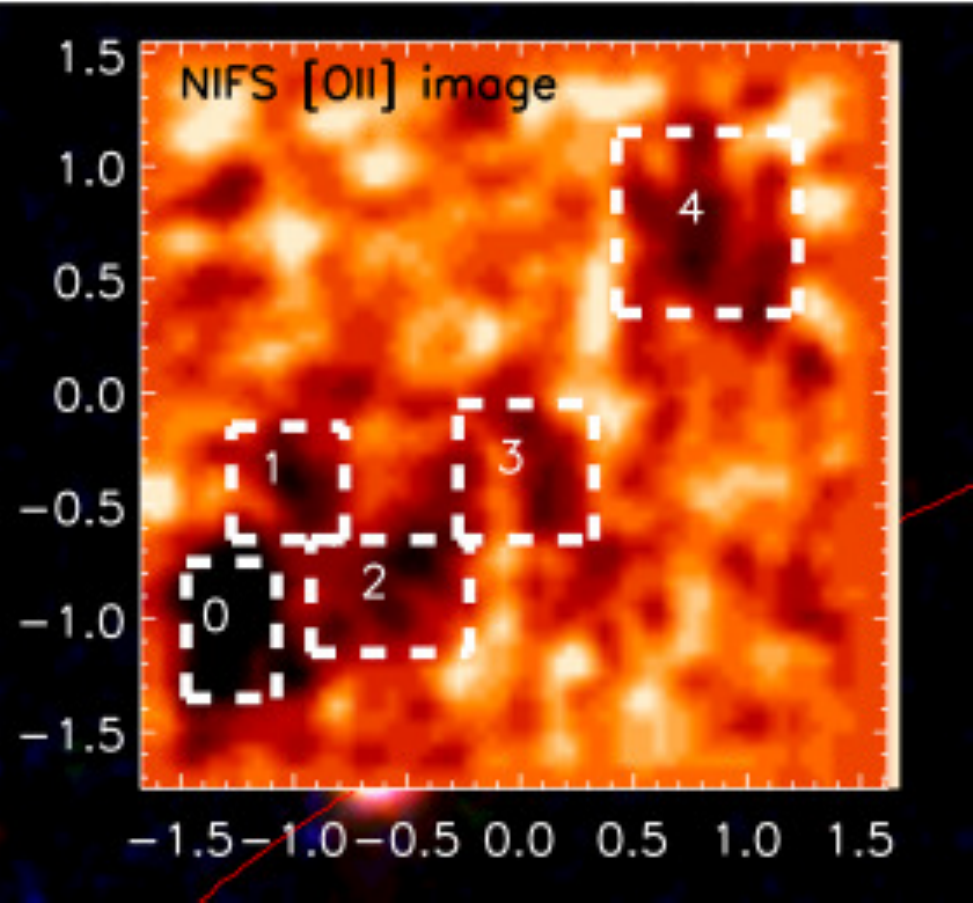}
\includegraphics[width=0.234\textwidth]{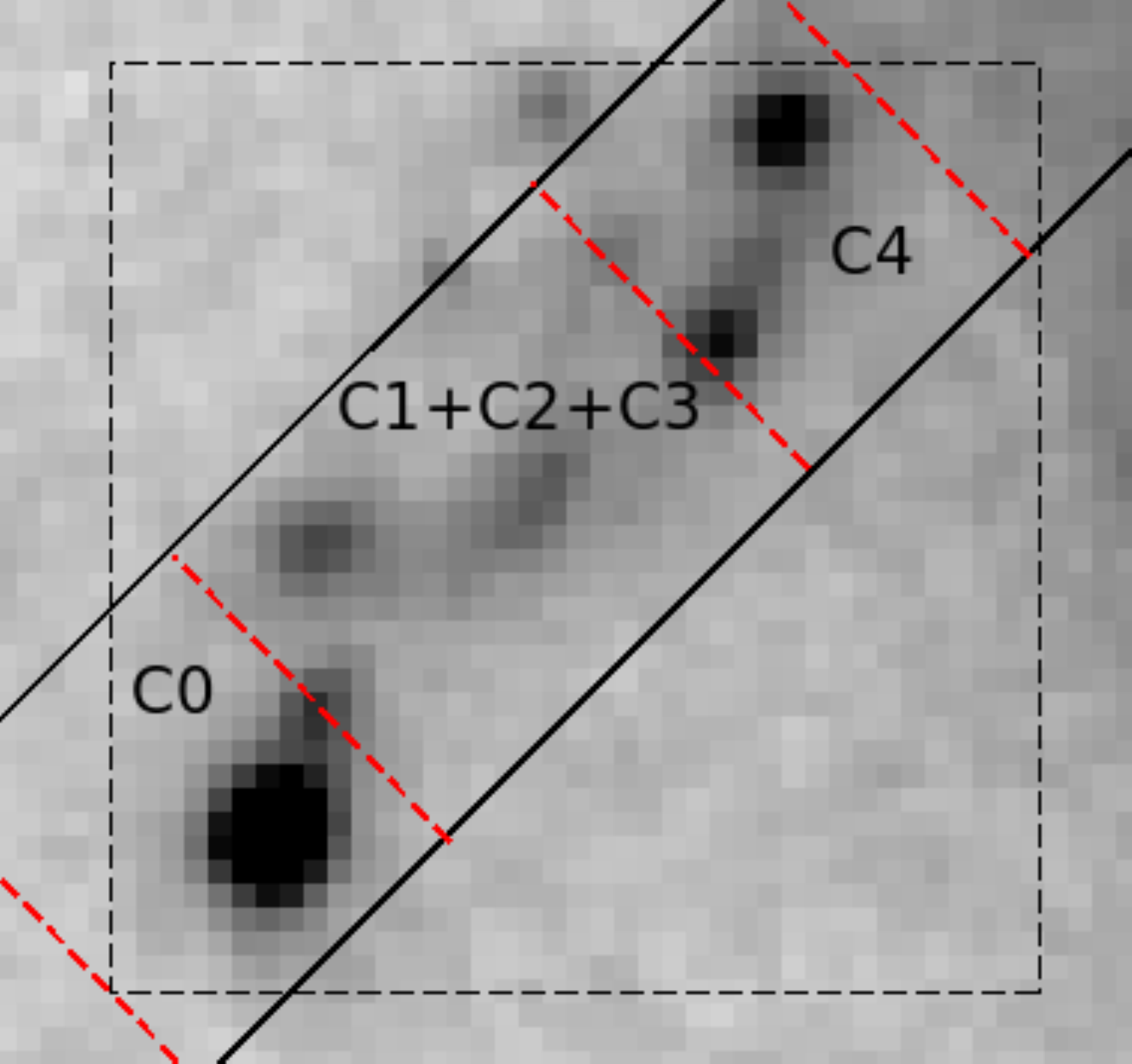}
\caption{Resolved spectroscopy in the highly-magnified source MS1358 at $z=4.93$. (Left) Distribution of [OII] emission from the integral field observations of \citet{Swinbank09}. (Right) HST ACS I-band image overlaid  with the 1 arcsec DEIMOS slit (solid black) and the \citet{Swinbank09} integral field image (dashed black). DEIMOS spectra were extracted for three regions containing clumps C0, C1+C2+C3 and C4
following the nomenclature of \citet{Swinbank09} (also marked in the left image).}
\label{fig:clump}
\end{figure}

Adopting the lensing magnification from our LTM model (\S3.3) and physical parameters for MS1358 calculated in \citet{Swinbank09}, its total stellar mass is $\sim7\pm2\times10^8$\msun and its integrated star-formation rate(SFR) is $\sim75\pm15$\msun yr$^{-1}$. The source itself is only a few kpc$^2$ in size. Since the individual clumps are conveniently linearly distributed along the arc, we have extracted spectra of each of 3 regions for specific clumps identified by \citet{Swinbank09}. Due to the limited spatial resolution of our ground-based spectroscopy and the signal to noise of our data, we group 3 of the smaller clumps (C1-3) together in extracting the spectra. The resulting  arrangement targeting 3 regions (C0, C1-3 and C4) is shown in Figure \ref{fig:clump} alongside the [OII] image derived from the \citet{Swinbank09} integral field unit observations.

\begin{figure*}
  \centering
  \includegraphics[width=0.6\textwidth]{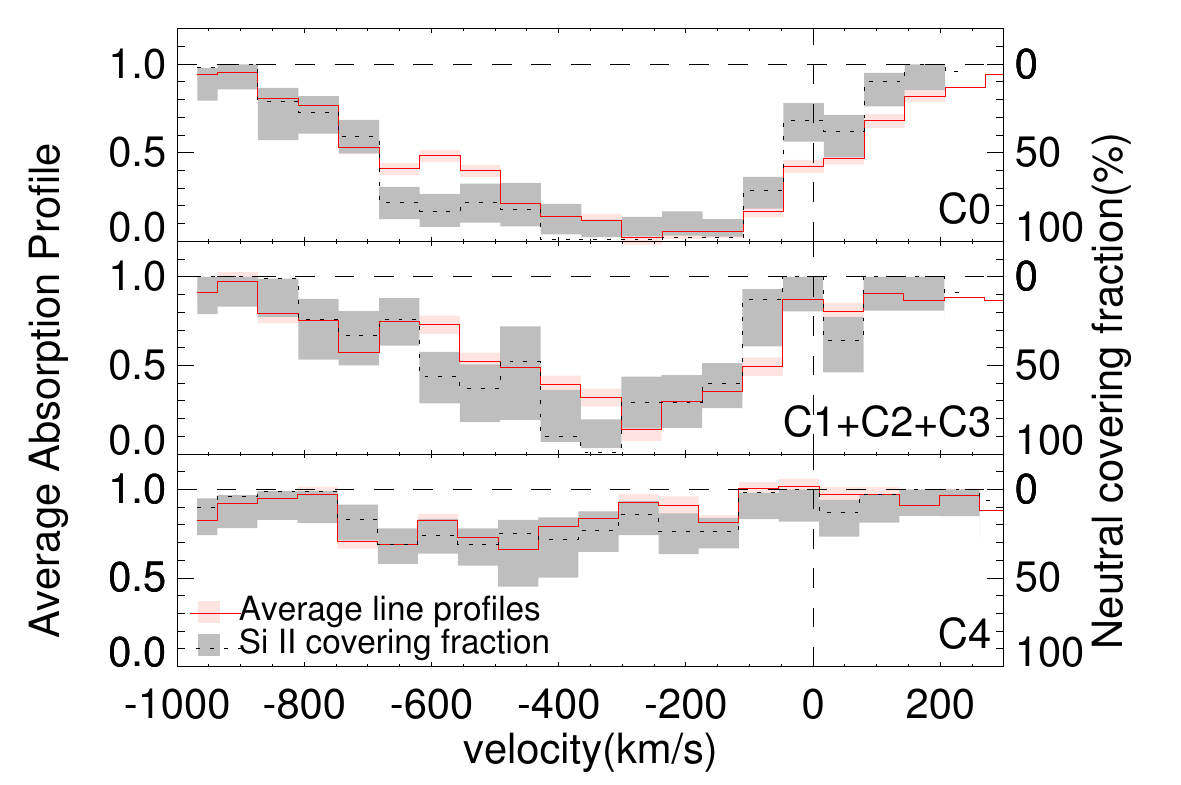}
  \caption{Average low-ionization absorption line profiles and covering fractions of \textit{individual clumps} in the highly-magnified source MS1358. The variation in the maximum absorption depths seen within the source is comparable to that seen across the integrated analysis of our enlarged sample in Figure \ref{fig:covfrac}.}
  \label{fig:covfrac_clumps}
\end{figure*}

Importantly, we find there is a significant variation in the low-ionization absorption profiles across the source (Figure \ref{fig:covfrac_clumps}). The inferred covering fraction ranges from near-complete ($100\%$) in C0 to only $\simeq40\%$ in C4. The variation within MS1358 is comparable to that across the integrated analysis of our full sample shown in Figure \ref{fig:covfrac}. As expected, it suggests that the escape fraction is 
governed by small-scale physics associated with star-forming regions. The variation between individual clumps indicates that covering fractions derived from integrated spectra may not be representative of the sum of all regions, and we can examine the degree of this effect. The SFR-weighted total of individual clumps is $f_{cov}\sim0.9$ whereas we find $f_{cov}\sim0.8$ based on the integrated light (Figure 7). It is possible that even higher covering fractions would be found at finer spatial resolution, although we note that \citet{Borthakur14} found $f_{esc,rel}\approx 1 - f_{cov}$ in a compact source of similar size to the clumps studied here ($\sim$ few hundred pc). We conclude that spatial variation in the covering fraction therefore reduces the limit on $f_{esc}$ by a factor of $\simeq 2$ in this source. To help determine the extent to which this source is representative of the larger sample, in the following sections we examine trends in our sample as a function both local and integrated physical properties.

\subsection{Variations with the Ly$\alpha$ Equivalent Width}

Since both Ly$\alpha$ and LyC radiation are scattered by clouds of neutral hydrogen, the low column density region from which LyC can escape should also provide an escape route for Ly$\alpha$ photons. A suite of radiative transfer simulations in clumpy ISM models by \citet{Dijkstra16} predicts a strong correlation between the escape fractions of \Lya and LyC. Indeed, observations indicate an anti-correlation between the low-ionization absorption depth and Ly$\alpha$ emission line equivalent width (W$_{Ly\alpha}$) \citep[e.g.][]{Jones13,Jones12,Erb14,Shapley03}. If this reflects the correlation between the \Lya and LyC escape fractions, not only does it imply that low-ionization absorption depths are a valuable proxy for the LyC escape fraction but it also provides observational support for the clumpy ISM models.     

\begin{figure}
  \centering
  \includegraphics[width=0.5\textwidth]{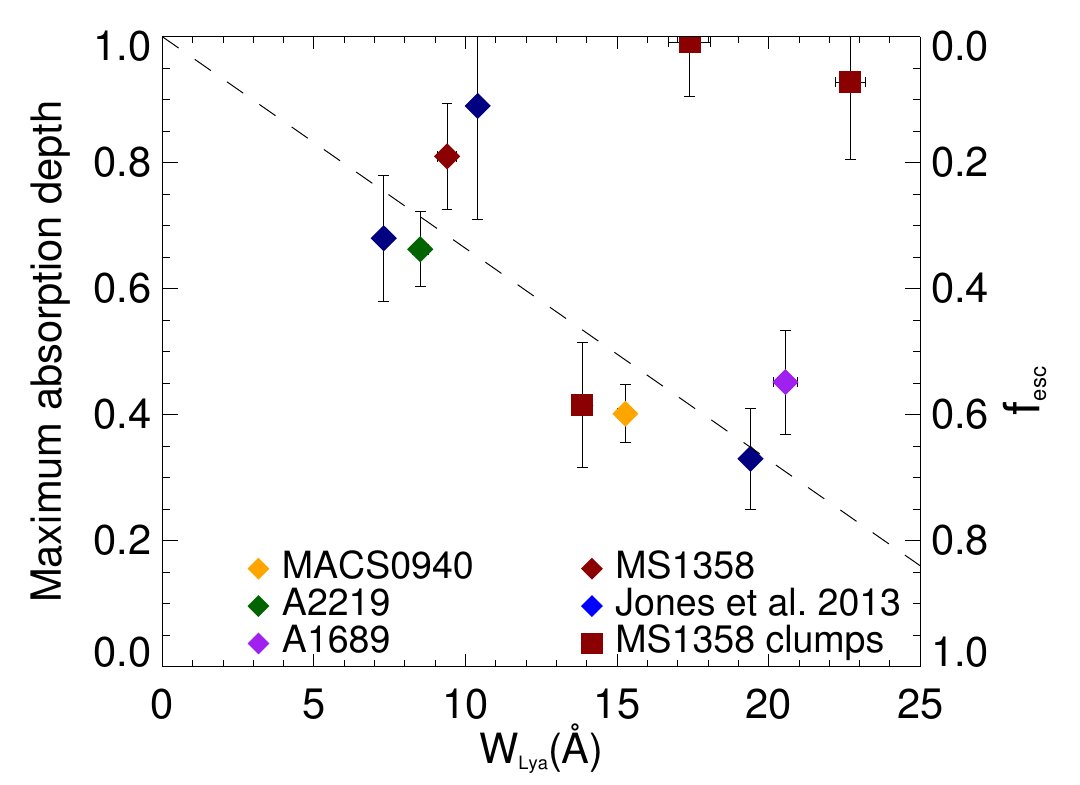}
  \caption{Correlation of the Ly$\alpha$ emission line equivalent width with the maximum low-ionization absorption depth, which supports the expected anti-correlation between the HI covering fraction and  \fesc. The dashed line show the earlier correlation published by \citet{Jones13}. With the exception of the intensely active clumps in MS1358, the new data supports the correlation. The scatter at high W$_{\textrm{Ly}\alpha}$ is consistent with that seen at high f$_{\textrm{esc,Ly}\alpha}$ found in simulations \citep{Dijkstra16}.} 
  \label{fig:lyaew}
\end{figure}

We plot the relation between W$_{Ly\alpha}$ and the low-ionization absorption depth in Figure \ref{fig:lyaew} combining both the integrated measures and those for the clumps identified in MS1358 (\S4.5).
Most of the new data points support the anti-correlation found in \citet{Jones13} as shown in dashed-line. However, two clumps within MS1358 (C0 and C1+C2+C3) appear as outliers although are consistent with the large scatter found in the \citet{Dijkstra16} simulations. The origin of the large scatter in the simulations is the cloud-covering factor (the average number of clouds along each sight-line around the region). There is a higher Poisson probability that a line of sight would be free from obscuration when the cloud-covering factor is low. If we interpret Figure \ref{fig:lyaew} accordingly to the simulation model, we can infer that the two clumps in MS1358 have a higher number of clouds given their increased star-formation rates.

\subsection{Variations with Reddening}

\cite{Reddy16} find that galaxies in their $z\simeq$3 sample with redder UV continua tend to have larger covering fractions of neutral hydrogen. They propose this correlation, modeled according to contributions from both photoelectric absorption and dust extinction, could form the basis of a new method to infer the escape fraction of high redshift galaxies, thereby bypassing the assumption of our method that the opacity of low ionization metal lines represent a reasonable proxy for that of neutral hydrogen. Although a correlation between the line of sight extinction, as inferred from the color excess $E(B-V)$, and $f_{cov}$ seems reasonable, for the integrated spectra for our full sample, we nonetheless found no strong dependence between our absolute covering fraction and $E(B-V)$. However, the three data points from the resolved data on MS1358 suggest that regions with higher covering fractions within the galaxy have redder UV continua and also stronger star formation rates as shown in Table \ref{table:SFR}. We will explore the variation of absorption line strength and star formation rates in the next section.

\subsection{Variations with the Star Formation Rate}

Our spatially-resolved observations of MS1358 presented in \S4.5 suggested that the implied escape fraction of ionizing photons depends sensitively on the small-scale structure of its star-forming regions. Here we consider the origin of these variations in terms of both the local and integrated star-formation rates.

The star formation rate (SFR) may have two opposing effects on the escape fraction. First, the SFR is directly correlated with the column density of gas, including LyC-absorbing HI as well as molecular H$_2$ and dust \citep{Krumholz12,Bigiel08,Bigiel10} However, strong feedback and radiation pressure in high SFR regions may clear the absorbing clouds resulting in a higher escape fraction. Understanding the balance between these two effects in early galaxies is critical in resolving the question of whether galaxies are capable of maintaining an ionized IGM. 

\begin{figure}
  \centering
  \includegraphics[width=0.5\textwidth]{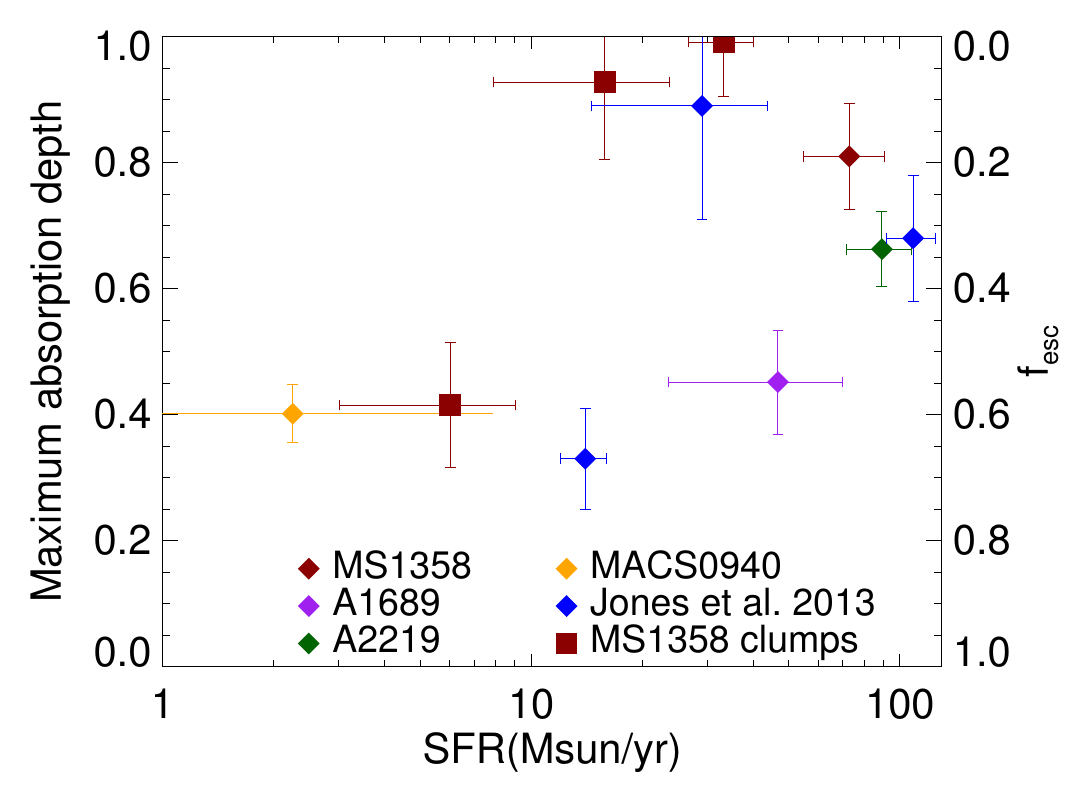}
  \caption{The correlation between the intrinsic star formation rate (after correction for the local or integrated magnification) and the maximum low-ionization absorption depth and covering fraction. Galaxies with high star formation rates tend to have higher covering fractions (and therefore lower value of \fesc). The trend is apparently stronger in individual star-forming clumps.}
  \label{fig:sfr_fc}
\end{figure}

According to our data shown in Figure \ref{fig:sfr_fc}, we see tentative evidence of a modest anti-correlation between SFR and the inferred escape fraction. Formally this trend is non-zero at the 2.8$\sigma$ level, with a correlation coefficient of $-0.53\pm0.19$ for the galaxy-integrated measurements. The trend is apparently stronger (though statistically consistent) for the individual clumps in MS1358 as compared to that seen for the integrated values across the galaxies in the sample. If real, this difference might be expected since the escape fraction inferred for an entire galaxy is necessarily a spectroscopic measure averaged over clumps with a range of both low and high SFRs. The observed anti-correlation can be interpreted in two independent ways. Firstly, regions of low SFR are associated with lower column densities of both interstellar and outflowing material (including HI and dust). Therefore, the probability of holes permitting leaking LyC photons is higher. Alternatively, regions with lower SFR may represent those whose rates were higher in the past and are now in decline. The associated delay time increases the chance that feedback effects can clear any covering clouds. Although both explanations may be valid, we consider the latter more likely because only a small column density of HI gas is needed to extinguish any leaking LyC radiation. Regardless of the surface density of low ionization gas, it seems feedback will be the more effective factor in governing both a low SFR and a smaller HI covering fraction. 

Most radiative transfer hydrodynamical simulations of the resolved ISM support our conjecture that a time delay will increase the escape fraction. Star formation may become increasingly `burst-like' at high redshift\citep{Stark09} and the escape fraction will then fluctuate on short time scales, out of phase with the SFR. \citet{Kimm14} find a delay of $\simeq$10 Myr for a covering cloud to be dispersed by feedback from supernovae, while \citet{Ma15} found a slightly shorter delay of $\simeq$3-10 Myr. A larger delay will increase the strength of the anti-correlation between SFR and escape fraction.

One cautionary remark is relevant in considering how to interpret Figure \ref{fig:sfr_fc} given the SFR values are inferred from gravitationally-lensed galaxies using our lensing model (\S 3.3). While the intrinsic SFR can be reliably estimated by dividing the observed value by the appropriate magnification, the covering fraction is measured from the absorption line depths in the image plane. Hence, there is a potential bias such that the covering fraction so measured will refer to that of the regions with the highest magnifications.  It would only be possible to appropriately transfer the absorption line spectra into the source plane with resolved spectroscopy using an integral field unit spectrograph. Such a bias would, however, only be significant if there is a significant magnification gradient along an arc, e.g.. when it is close to a critical curve. In our sample, this issue is not important. Both A1689 and A2219 have low and uniform magnifications and the individual MS1358 clumps are sufficiently small as to have uniform magnifications within each region. When MS1358 is considered as an entirety, we excluded arc B (Figure \ref{fig:slit}) which has a high variation in magnification. MACS0940 is the only system where the variation in magnification is significant. However, its SFR is relatively low so such a bias is unlikely to affect its location on the relation in Figure \ref{fig:sfr_fc}.

\subsection{Redshift Evolution and Reonization}

It is clear that, despite the significant uncertainties, our moderate covering fractions imply absolute escape fractions that are higher than the averages observed via direct measures of the LyC radiation at lower redshift \citep[e.g.][]{Mostardi15}. On the other hand, if star formation is increasingly burst-like at high redshift where galaxies are physically more compact, it is quite likely from the arguments discussed above, that there is increase in \fesc\ with redshift. Theoretical studies present conflicting predictions on the question of evolution in \fesc. \citet{Ferrara13} present an analytical calculation that illustrates why the average escape fraction should decrease with time during the reionization era. With an assumed value for the efficiency of star-formation, they estimate \fesc\ decreased by a factor of 3 from $z=10$ to $z=6$. However, sample-limited radiative hydrodynamical simulations show no significant variation over the same redshift interval \citep{Ma15}.

\begin{figure}
  \centering
  \includegraphics[width=0.5\textwidth]{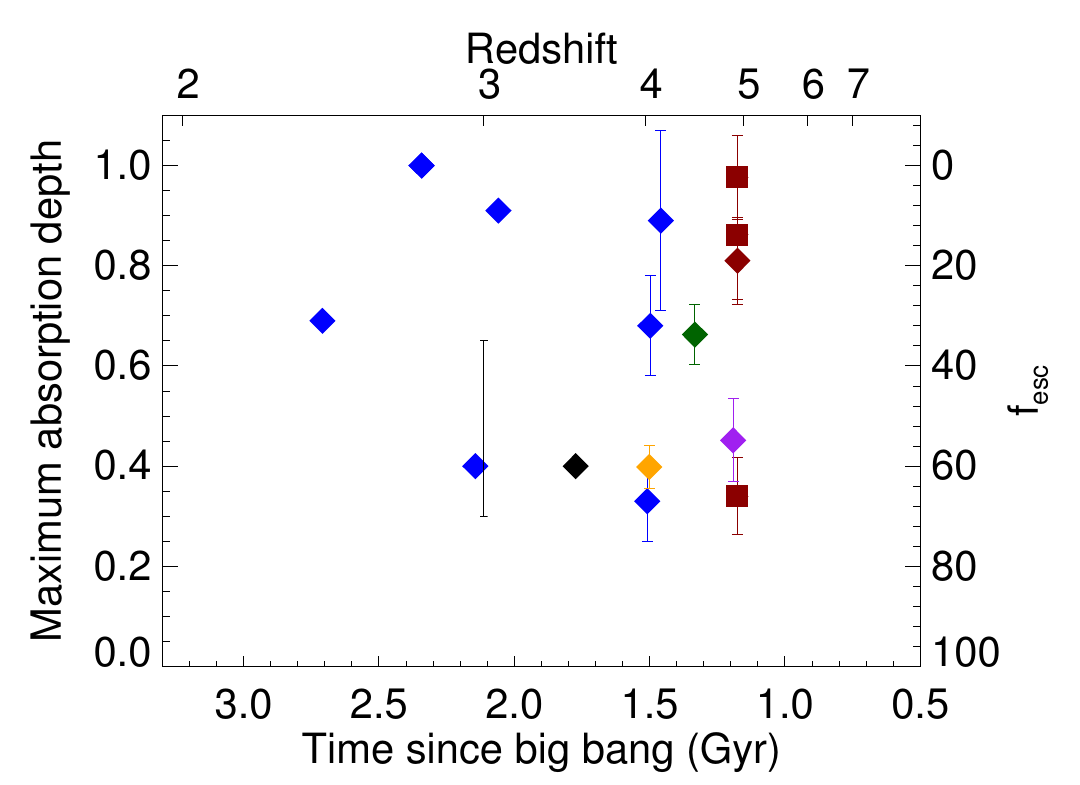}
  \caption{Time-dependence of the maximum low-ionization absorption depth with cosmic time for the present and earlier samples from \citet{Jones13} and \citet{Reddy16} spanning the redshift range $2.5<z<5$. No trend is apparent and the scatter supports our deduction that the escape fraction is governed by the small scale behavior of star-forming regions which fluctuate in their activities over short timescales. The color scheme follows that in Figure \ref{fig:sfr_fc}. The black error bar and the black data point show $f_{cov,obs}$ derived from low-ionization absorption profiles, in a stacked spectrum of $z\sim3$ galaxies from \citet{Reddy16} and in a spectrum of a lensed $z=3.5$ galaxy from \citet{Patricio16} respectively.}
  \label{fig:z_fc}
\end{figure}

In compiling the data for Figure \ref{fig:z_fc} we include observations at redshifts $z<4$ from \citet{Jones13} (not shown in Table 1), \citet{Reddy16}, and \citet{Patricio16}. Across this increased redshift range, the figure shows that we cannot yet detect any significant time dependence. This is because the scatter in covering fraction is dominant and apparently constant over time, in agreement with our deduction that the escape fraction depends largely on the small-scale structure of clouds, their physical association with star formation regions and the short timescale of activity, and consistent with the behavior of simulated galaxies in \citet{Ma15}. 

As shown in Figure \ref{fig:z_fc}, absorption line measurements of comparable quality to those presented here are still scarce at $z=2$-3. Recent studies of the ``cosmic horseshoe" at $z=2.4$ further highlights the challenge in comparing direct and indirect methods. The covering fraction was found to be $\sim60$\% using methods equivalent to ours \citep{Quider09}, yet imaging below the Lyman limit indicates $f_{esc,rel}<0.08$ \citep{Vasei16}. This might be reconciled if $f_{cov}$ overestimates $f_{esc}$ by a factor of 5, or by low IGM transmission ($\sim$20\% probability, \citet{Vasei16}, or a combination of both effects. At present the cause is not clear and it will be of great interest to examine low-ion covering fractions in more low redshift sources with direct $f_{esc}$ measurements. This is becoming feasible following recent LyC detections in nearby galaxies using HST/COS (e.g. \citet{Izotov16, Leitherer16}). \citet{Reddy16} present new measures of the covering fraction of both the low ionization gas and neutral hydrogen for stacked spectra derived from their Keck survey at $z\simeq$3 providing a valuable complementary approach to that discussed here. Although we have highlighted the possible difficulties of inferring a mean covering fraction from a stack of many spectra with diverse kinematic absorption line profiles (such as the variety shown in Figure 4), \citet{Reddy16} draw from a much larger sample and have questioned some of the assumptions inherent in the use of low ionization gas alone. However, within the observational uncertainties at this stage, both approaches give reasonably consistent results for the inferred escape fraction at $z>3$.

Despite significant observational and theoretical effort, direct evidence for a rising escape fraction at high redshift is still lacking due to the scarcity of LyC detections. Our data set suggests a median $f_{esc,abs}=19\pm6$\% at $z=4$-5 if we interpret the covering fraction as $f_{cov}=1-f_{esc,rel}$. However, spatial variation may lower this value to $f_{esc,abs}\approx10$\% if the resolved properties of MS1358 are typical of our sample, and other effects discussed in Section 4.4 may reduce the value slightly further (to $\sim7$\%). These estimates are certainly compatible with independent determinations from the IGM, as well as an increase over the limits at $z\sim2$. Furthermore the demographic trends in Figures 7 and 8 indicate that we may expect escape fractions to increase with redshift, commensurate with increasing $W_{Lya}$ \citep[e.g.][]{Law12}. We speculate that a key difference may be that star formation is more burst-like (e.g. higher specific SFR; \citet{Schreiber15} at higher $z$. Intense short-duration starburst episodes may be capable of clearing sightlines through the local ISM resulting in increased ionizing escape during periods of lower activity, giving rise to the trend in Figure 8.

\section{Summary}

We present intermediate resolution absorption line spectroscopy for four new gravitationally-lensed sources which, together with our earlier work \citep{Jones13}, provides a total sample of seven star-forming galaxies in the redshift interval $4<z<5$. Our signal to noise in all cases is adequate to sample the depth of absorption in low ionization species which can act as a valuable indicator of the covering fraction of neutral hydrogen.  We describe two methods to convert this covering fraction into estimates of the fraction of Lyman continuum photons that are capable of reaching the intergalactic medium and undertake a critical analysis of possible uncertainties in our results. Our enlarged sample strengthens the trends discovered in our earlier work. In particular we find:

\begin{itemize}

\item{} The amount of absorbing gas varies significantly from galaxy to galaxy across our sample. However, only two systems are consistent with a complete coverage of low ionization gas. The median observed covering fraction is 66\%. Following a correction for reddening based on a dust-in-cloud model, we infer a median absolute escape fraction of Lyman continuum photons of $19\pm6$\%

\item{} We discuss various uncertainties in this estimate including the possibility that the stellar radiation has different covering fractions depending on its ionizing capability as well as scattered photons which could fill in the absorption line profiles mimicking a reduced covering fraction. We demonstrate that, even in the extreme case, these issues would only lower the escape fraction to no less than 11\%, a value in broad agreement with independent estimates based on recombination studies of the IGM probed by QSO absorption lines. Further study of low-ion absorption lines in nearby galaxies with recent LyC detections will help determine the degree to which covering fractions may overestimate $f_{esc}$.

\item{} For one of our lensed sources, MS1358, we demonstrate that the low ionization gas is clearly spatially inhomogeneous with covering fractions ranging from 40-100\% across various star-forming clumps. Although our sample generally strengthens the correlation we found earlier whereby the escape fraction increases with the equivalent width of Ly$\alpha$ emission, the scatter increases when we consider trends in the individual star-forming clumps in MS1358, consistent with recent numerical simulations. As the escape fraction is anti-correlated with the local star formation rate, this suggests a time delay between a burst and the emergence of leaking Lyman continuum photons. 

\item{} Overall, our observations indicate significant variations in the escape fraction, both from galaxy-to-galaxy and spatially within an individual galaxy. This pattern is consistent with the ability of escape being governed by the small scale behavior of star-forming regions which fluctuate in their  activities over short timescales. This conclusion supports the suggestion that the escape fraction may increase toward the reionization era when star formation becomes more energetic and burst-like in nature.

\end{itemize}

\section{Acknowledgements}

We acknowledge useful discussions with Phil Hopkins, Evan Kirby, Xiangcheng Ma, Brant Robertson, Brian Siana and Alice Shapley.
TAJ acknowledges support from NASA through Hubble Fellowship grant HST-HF2-51359.001-A awarded by the Space Telescope Science Institute, which is operated by the Association of Universities for Research in Astronomy, Inc., for NASA, under contract NAS 5-26555.
RSE acknowledges support from the European Research Council through an Advanced Grant FP7/669253.
DPS acknowledges support from the National Science Foundation  through the grant AST-1410155.
AZ is supported by NASA through Hubble Fellowship grant \#HST-HF2-51334.001-A awarded by STScI, which is operated by the Association of Universities for Research in Astronomy, Inc. under NASA contract NAS~5-26555.
Data presented herein were obtained at the W. M. Keck Observatory, which is operated as a scientific 
partnership among the California Institute of Technology. the University of California and NASA. 
The Observatory was made possible by the generous financial support of the W. M. Keck Foundation.
The authors acknowledge the very significant cultural role that the
summit of Mauna Kea has always had within the indigenous Hawaiian community.
We are most fortunate to have the opportunity to conduct observations from this mountain.

\end{document}